\documentstyle[aps,epsf,preprint]{revtex}
\tightenlines
\begin{document}
\draft
\title{Analysis of the nature of the $\phi\to\gamma\pi\eta$ and
 $\phi\to\gamma\pi^0\pi^0$ decays.}
\author
{N.N. Achasov
\thanks{achasov@math.nsc.ru}
 and V.V. Gubin
\thanks{gubin@math.nsc.ru}
  \\ Laboratory of Theoretical Physics,
\\ Sobolev Institute for Mathematics \\ 630090, Novosibirsk-90,
Russia}
\date{\today}
\maketitle
\begin{abstract}
 We study interference patterns in the
$\phi\to(\gamma a_0+\pi^0\rho)\to\gamma\pi\eta$ and
 $\phi\to(\gamma f_0+\pi^0\rho)\to\gamma \pi^0\pi^0$ reactions.
 Taking into account the interference,  we fit the
 experimental data and show that the background reaction
 does not distort the $\pi^0\eta$ spectrum in the decay
$\phi\to\gamma\pi\eta$ everywhere over the energy region
 and does not
 distort the $\pi^0\pi^0$ spectrum in the decay
$\phi\to\gamma\pi^0\pi^0$ in the wide region of the $\pi^0\pi^0$
system invariant mass, $m_{\pi\pi}>670$ MeV, or when the photon
energy is less than  300 MeV. We discuss the details of the scalar
meson production in the radiative decays and note that there are
reasonable arguments in favor of the one-loop mechanism  $\phi\to
K^+K^-\to\gamma a_0$ and $\phi\to K^+K^-\to\gamma f_0$. We discuss
also distinctions between the four-quark, molecular, and two-quark
models and argue that the Novosibirsk data give evidence  in favor
of the four-quark  nature of the scalar $a_0(980)$ and $f_0(980)$
mesons.
\end{abstract}


\section{Introduction}

As  was shown in a number of papers, see Refs.
\cite{achasov-89,close-93,nutral,shevchenko,lucio,phase} and
references therein, the study of the radiative decays
$\phi\to\gamma a_0\to\gamma\pi\eta$ and $\phi\to\gamma f_0\to
\gamma\pi\pi$ can shed light on the problem of the scalar
$a_0(980)$ and $f_0(980)$ mesons. These decays have been studied
not only theoretically but also experimentally. Present time data
have already been obtained from Novosibirsk with the detectors SND
 \cite{snd-1,snd-2,snd-fit,snd-ivan} and CMD-2 \cite{cmd},
 which give the following
branching ratios :
$BR(\phi\to\gamma\pi\eta)=(0.88\pm0.14\pm0.09)\cdot10^{-4}$
\cite{snd-fit}, $BR(\phi\to\gamma\pi^0\pi^0)=
(1.221\pm0.098\pm0.061)\cdot10^{-4}$ \cite{snd-ivan} and
$BR(\phi\to\gamma\pi\eta)=(0.9\pm0.24\pm0.1)\cdot10^{-4}$,
$BR(\phi\to\gamma\pi^0\pi^0)=(0.92\pm0.08\pm0.06)\cdot10^{-4}$
\cite{cmd}.

 These data give evidence in favor of the four-quark $(q^2\bar
q^2)$ \cite{achasov-89,jaffe,ach-84,ach-91,ach-98,black} nature of
the scalar $a_0(980)$ and $f_0(980)$ mesons. Note that the
isovector $a_0(980)$ meson is produced in the radiative $\phi$
meson decay
 as intensively as the well-studied $\eta'$ meson involving
 essentially strange quarks $s\bar s$
($\approx66\%$), responsible for the decay.

As shown in Refs. \cite{achasov-89,nutral,bramon}, the background
situation for studying the radiative decays $\phi\to\gamma
a_0\to\gamma\pi^0\eta$ and $\phi\to\gamma f_0\to \gamma\pi^0\pi^0$
is very good. For example, in the case of the decay $\phi\to\gamma
a_0\to\gamma\pi^0\eta$, the process
$\phi\to\pi^0\rho\to\gamma\pi^0\eta$ is the dominant background.
The estimation for the soft, by strong interaction standard,
photon energy, $\omega<100$ MeV, gives
$BR(\phi\to\pi^0\rho^0\to\gamma\pi^0\eta,\omega<100\ \mbox{MeV})
\approx 1.5\cdot10^{-6}$. The influence of the background process
is negligible, provided
 $BR(\phi\to\gamma a_0\to\gamma\pi^0\eta,\omega<100\ \mbox{MeV})\geq10^{-5}$. In
this paper, in Sec. II, we calculate the expression for the
$\phi\to\gamma\pi^0\eta$ decay amplitude taking into account the
interference between the $\phi\to\gamma a_0\to\gamma\pi^0\eta$ and
$\phi\to\pi^0\rho\to\gamma\pi^0\eta$ processes. We show that for
the obtained experimental data the influence of the background
processes is negligible everywhere over the photon energy region.

The situation with  $\phi\to\gamma f_0\to \gamma
 \pi^0\pi^0$ decay is not much different. As was shown in
\cite{achasov-89,nutral,bramon} the dominant background is the
$\phi\to\pi^0\rho^0\to\gamma\pi^0\pi^0$ process with
$BR(\phi\to\pi^0\rho^0\to\gamma\pi^0\pi^0,\omega<100\ \mbox{MeV})
\approx 6.4\cdot10^{-7}$. The influence of this background process
is negligible, provided $BR(\phi\to\gamma
f_0\to\gamma\pi^0\pi^0,\omega<100\ \mbox{MeV})\geq5\cdot10^{-6}$.

 The exact calculation of the interference patterns between the decays
$\phi\to\gamma f_0\to \gamma\pi^0\pi^0$ and
$\phi\to\rho^0\pi\to\gamma\pi^0\pi^0$, which we present in this
paper in Sec. III, shows that the influence of the background in
the decay
 $\phi\to\gamma\pi^0\pi^0$ for the obtained experimental data
 is negligible in the wide region of
 the $\pi^0\pi^0$ invariant mass, $m_{\pi\pi}>670$ MeV,
 or in the photon energy region $\omega<300$ MeV.

 In Sec. IV we discuss the mechanism of the scalar meson production in the
 radiative decays and show that experimental data obtained in Novosibirsk
give the reasonable arguments in favor of the one-loop mechanism
$\phi\to K^+K^-\to\gamma a_0$ and $\phi\to K^+K^-\to\gamma f_0$ of
these  decays . In the same place we discuss also distinctions
between the four-quark, molecular, and two-quark models and
explain why these data give evidence
 in favor of the four-quark  nature of the
scalar $a_0(980)$ and $f_0(980)$ mesons.

\section{Interference between the reactions \lowercase{$\phi\to\gamma
a_0\to\gamma \pi^0\eta$} and
\lowercase{$\phi\to\pi^0\rho^0\to\gamma\pi^0\eta$}.}

As was shown in Refs.  \cite{achasov-89,nutral} the background
process $e^+e^-\to\phi\to\pi^0\rho^0\to\gamma\pi^0\eta$ is
dominant. The amplitudes of the processes
$e^+e^-\to\rho^0(\omega)\to\eta\rho^0(\omega)\to\gamma\pi^0\eta$
are much less than the amplitudes of the
 $e^+e^-\to\rho^0(\omega)\to\pi^0\omega(\rho^0)\to\gamma\pi^0\eta$
 processes.
 In its turn, the amplitudes of the
$e^+e^-\to\rho^0(\omega)\to\pi^0\omega(\rho^0)\to\gamma\pi^0\eta$
processes are much less than the amplitudes of the
$e^+e^-\to\phi\to\pi^0\rho^0\to\gamma\pi^0\eta$  processes. The
amplitude of the $e^+e^-\to\phi\to\eta\phi\to\gamma\pi^0\eta$
process is also much less  than the amplitude of
$e^+e^-\to\phi\to\pi^0\rho^0\to\gamma\pi^0\eta$ process.

The amplitude of the background process
$\phi(p)\to\pi^0\rho^0\to\gamma(q)\pi^0(k_1)\eta(k_2)$ is
\begin{equation}
M_B=\frac{g_{\phi\rho\pi}g_{\rho\eta\gamma}}{D_{\rho}(p-k_1)}
\phi_{\alpha}k_{1\mu}p_{\nu}\epsilon_{\delta}(p-k_1)_{\omega}q_{\epsilon}
\epsilon_{\alpha\beta\mu\nu}\epsilon_{\beta\delta\omega\epsilon}.
\end{equation}

For the amplitude of the signal $\phi\to\gamma
a_0\to\gamma\pi^0\eta$ we use the model suggested in Ref.
\cite{achasov-89}, in which the one-loop mechanism of the decay
$\phi\to K^+K^-\to\gamma a_0$ is considered:
\begin{equation}
M_a=g(m)\frac{g_{a_0K^+K^-}g_{a_0\pi\eta}}{D_{a_0}(m)}\left
((\phi\epsilon)- \frac{(\phi q)(\epsilon p)}{(pq)}\right )
\label{a0signal}\,,
\end{equation}
where  $m^2=(k_1+k_2)^2$, $\phi_{\alpha}$ and $\epsilon_{\mu}$ are
the polarization vectors of $\phi$ meson and photon, the function
$g(m)$ is determined
 in Refs. \cite{achasov-89,nutral}. The mass spectrum is
\begin{equation}
\frac{\Gamma(\phi\to\gamma\pi\eta)}{dm}=\frac{d\Gamma_{a_0}(m)}{dm}+
\frac{d\Gamma_{back}(m)}{dm}\pm \frac{d\Gamma_{int}(m)}{dm}\,,
\end{equation}
where the mass spectrum for the signal is
\begin{eqnarray}
\frac{d\Gamma_{a_0}(m)}{dm}=\frac{2}{\pi}\frac{m^2\Gamma(\phi\to\gamma
a_0(m))\Gamma(a_0(m)\to\pi\eta)}{|D_{a_0}(m)|^2}=
\frac{2|g(m)|^2p_{\eta\pi}(m_{\phi}^2-m^2)}
{3(4\pi)^3m_{\phi}^3}\left
|\frac{g_{a_0K^+K^-}g_{a_0\pi\eta}}{D_{a_0}(m)}\right |^2\,.
\label{spectruma0}
\end{eqnarray}

Accordingly, the mass spectrum for  the background process
$e^+e^-\to\phi\to\pi^0\rho\to\gamma\pi^0\eta$  is

\begin{equation}
\frac{d\Gamma_{back}(m)}{dm}=\frac{(m_{\phi}^2-m^2)p_{\pi\eta}
}{128\pi^3m_{\phi}^3}\int_{-1}^{1}dxA_{back}(m,x)\,,
\end{equation}
where
\begin{eqnarray}
&&A_{back}(m,x)=\frac{1}{3}\sum|M_B|^2= \nonumber \\
&&=\frac{1}{24}\left (m_{\eta}^4m_{\pi}^4+2m^2m_{\eta}^2m_{\pi}^2
\tilde{m_{\rho}}^2-2m_{\eta}^4m_{\pi}^2\tilde{m_{\rho}}^2-2m_{\eta}^2m_{\pi}^4
\tilde{m_{\rho}}^2+\right. \nonumber \\ && \left.
2m^4\tilde{m_{\rho}}^4-
2m^2m_{\eta}^2\tilde{m_{\rho}}^4+2m_{\eta}^4\tilde{m_{\rho}}^4
-2m^2m_{\pi}^2\tilde{m_{\rho}}^4+4m_{\eta}^2m_{\pi}^2\tilde{m_{\rho}}^4
+m_{\pi}^4\tilde{m_{\rho}}^4+\right. \nonumber \\
&&\left. 2m^2\tilde{m_{\rho}}^6-
2m_{\eta}^2\tilde{m_{\rho}}^6-2m_{\pi}^2\tilde{m_{\rho}}^6+
\tilde{m_{\rho}}^8-2m_{\eta}^4m_{\pi}^2m_{\phi}^2-
2m^2m_{\eta}^2m_{\phi}^2\tilde{m_{\rho}}^2+\right. \nonumber \\
&&\left. 2m_{\eta}^2m_{\pi}^2m_{\phi}^2\tilde{m_{\rho}}^2-
2m^2m_{\phi}^2\tilde{m_{\rho}}^4+
2m_{\eta}^2m_{\phi}^2\tilde{m_{\rho}}^4-
2m_{\phi}^2\tilde{m_{\rho}}^6+ m_{\eta}^4m_{\phi}^4+
m_{\phi}^4\tilde{m_{\rho}}^4\right )\times\nonumber \\
&&\left
|\frac{g_{\phi\rho\pi}g_{\rho\eta\gamma}}{D_{\rho}(\tilde{m_{\rho}})}\right
|^2\,,
\end{eqnarray}
and
\begin{eqnarray}
&&\tilde{m_{\rho}}^2=m_{\eta}^2+\frac{(m^2+m_{\eta}^2-m_{\pi}^2)(m_{\phi}^2-
m^2)}{2m^2}-\frac{(m_{\phi}^2-m^2)x}{m}p_{\pi\eta}\nonumber \\
&&p_{\pi\eta}=\frac{\sqrt{(m^2-(m_{\eta}-m_{\pi})^2)
(m^2-(m_{\eta}+m_{\pi})^2)}}{2m}\,.
\end{eqnarray}

 The interference between the background process amplitude and the
 signal amplitude is written in the following way:

\begin{equation}
\frac{d\Gamma_{int}(m)}{dm}=\frac{(m_{\phi}^2-m^2)p_{\pi\eta}
}{128\pi^3m_{\phi}^3} \int_{-1}^{1}dxA_{int}(m,x)\,,
\end{equation}
where
\begin{eqnarray}
&&A_{int}(m,x)=\frac{2}{3}Re\sum M_aM_B^*=
\frac{1}{3}\left((m^2-m_{\phi}^2)\tilde{m_{\rho}}^2+
\frac{m_{\phi}^2(\tilde{m_{\rho}}^2-m_{\eta}^2)^2}{m_{\phi}^2-m^2}\right)
\times\nonumber\\
&&Re\left
\{\frac{g(m)g_{a_0K^+K^-}g_{a_0\pi\eta}g_{\phi\rho\pi}g_{\rho\eta\gamma}}
{D^*_{\rho}(\tilde{m_{\rho}})D_{a_0}(m)}\right\}\,.
\end{eqnarray}

The inverse propagator of  $a_0$ meson, $D_{a_0}(m)$, is presented
in Refs. \cite{achasov-89,nutral}. The inverse propagator of
 $\rho$ meson has the following expression
\begin{equation}
D_{\rho}(m)=m_{\rho}^2-m^2-im^2\frac{g^2_{\rho\pi\pi}}{48\pi}
\left (1-\frac{4m_{\pi}^2}{m^2}\right )^{3/2}\,.
\end{equation}

We use the coupling constant  $g_{\phi K^+K^-}=4.68\pm0.05$
obtained form the decay $\phi\to K^+K^-$ \cite{pdg}, and the
coupling constant $g_{\rho\eta\gamma}=0.572\pm0.08$ GeV$^{-1}$
obtained from the decay $\rho\to\eta\gamma$ \cite{dolinsky}, with
the help of the following expressions
\begin{equation}
\Gamma(\phi\to K^+K^-)=\frac{g_{\phi K^+K^-}^2}{48\pi}m_{\phi}
\left (1-\frac{4m_K^2}{m_{\phi}^2}\right )^{3/2},\ \ \
\Gamma(\rho\to\eta\gamma)=\frac{g_{\rho\eta\gamma}^2}{96\pi
m_{\rho}^3} \left (m_{\rho}^2-m_{\eta}^2\right )^3\,.
\end{equation}
The coupling constant $g_{\phi\rho\pi}=0.811\pm0.081$ GeV$^{-1}$
is obtained using the data on the decay
$\phi\to\rho\pi\to\pi^+\pi^-\pi^0$ \cite{pdg} with the help of the
formulas from the paper \cite{achasov-kozhevnikov}.

The fit of the experimental data from the SND detector
\cite{snd-fit},
  taking into account
 the relation  $g_{a_0\pi\eta}=0.85 g_{a_0K^+K^-}$ resulting from
 the $q^2\bar q^2$ model \cite{achasov-89}, chooses the
 constructive interference and gives

\begin{eqnarray}
&&m_{a_0}=985.51\pm0.8\ \ \mbox{MeV} \nonumber \\
&&g_{a_0K^+K^-}=2.747\pm0.428\ \ \mbox{GeV};\
\frac{g_{a_0K^+K^-}^2}{4\pi}=0.6\pm0.015\ \ \mbox{GeV}^2
\nonumber\\ &&\chi^2/dof=3.1/6 \label{parama0}\,.
\end{eqnarray}

\begin{figure}
\centerline{\epsfxsize=12cm \epsfysize=8cm \epsfbox{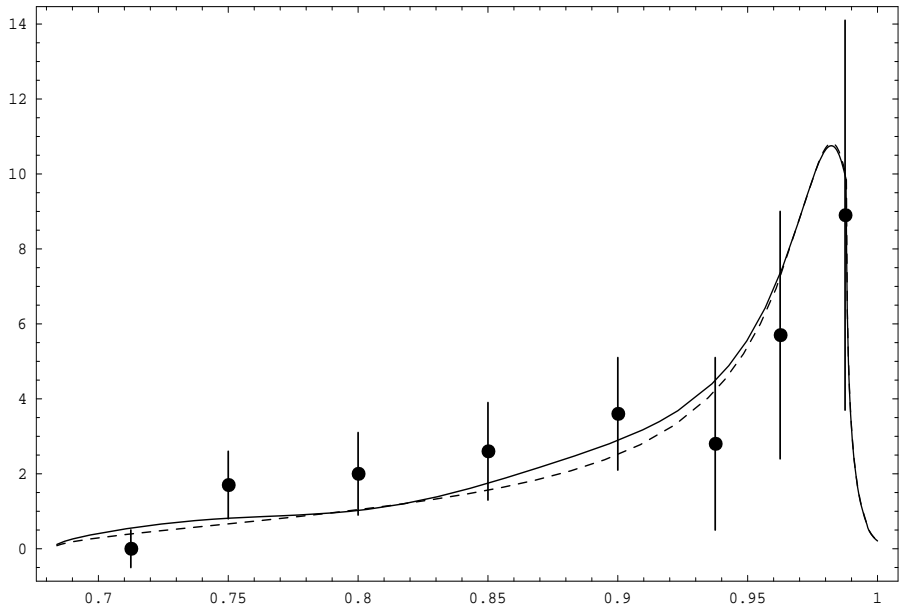}}
 \caption{ Fitting of $dBR(\phi\to\gamma\pi\eta)/dm\times 10^4\mbox{GeV}^{-1}$  with
the background is shown with the solid line, the signal
contribution is shown with the dashed line.
  } \label{figa0}
\end{figure}

The total branching ratio, taking into account the interference,
is $BR(\phi\to(\gamma
a_0+\pi^0\rho)\to\gamma\pi\eta)=(0.79\pm0.2)\cdot10^{-4}$, the
branching ratio of the signal is  $BR(\phi\to\gamma
a_0\to\gamma\pi\eta)=(0.75\pm0.2)\cdot10^{-4}$ and the branching
ratio of the background is
$BR(\phi\to\rho^0\pi^0\to\gamma\pi^0\eta)=3.43\cdot10^{-6}$. So,
the integral part of the interference is negligible. The influence
of the interference on the mass spectrum of the $\pi\eta$ system
is also negligible, see Fig. \ref{figa0}.

 The difference of the obtained parameters (\ref{parama0}) from the
  parameters found in
\cite{snd-fit}, which are $m_{a_0}=994\pm^{33}_{8}$ MeV,
$g_{a_0K^+K^-}^2/4\pi=1.05\pm^{0.36}_{0.25}$ GeV$^2$, is due to
the fact that in \cite{snd-fit} a more refined fitting was
performed considering  the event distribution inside of the each
bin. Notice that this difference is less than two standard
deviations.

Let us specially  emphasize that the value
$g_{a_0K^+K^-}^2/4\pi=0.6\pm0.015$ GeV$^2$ obtained by us  on no
account points to the possibility of the $K\bar K$ molecule
description \cite{close-93} of the $a_0$ meson. In the $K\bar K$
molecule model, the imaginary part of the $K^+K^-$ loop is
dominant because the real part of the $K^+K^-$ loop is suppressed
by the wave function of the molecule \cite{shevchenko}, see also
Sec. IV. Due to this fact, we have $BR(\phi\to\gamma
a_0\to\gamma\pi\eta)\approx1.5\cdot10^{-5}$ \cite{shevchenko} in
the $K\bar K$ molecule model at the same coupling constant and
$m_{a_0}=985$ MeV, which is almost by six times less than the
experimental value
$BR(\phi\to\gamma\pi\eta)=(0.88\pm0.14\pm0.09)\cdot10^{-4}$
\cite{snd-fit}. The divergence is by five standard deviations!
Besides, in the case of molecule, the bump in the spectrum of the
$\pi\eta$ system is much narrower  than the experimentally
observed, \cite{shevchenko}, see also Sec. IV.

\section{Interference between the
\lowercase{$e^+e^-\to\gamma f_0\to\gamma \pi^0\pi^0$} and
\lowercase{$e^+e^-\to\phi\to\pi^0\rho\to\gamma\pi^0\pi^0$}
reactions}

When analyzing the $\phi\to\gamma f_0\to\gamma\pi^0\pi^0$ decay,
one should take into account the mixing of the  $f_0$ meson with
the isosinglet scalar states. The whole formalism of the mixing of
two scalar $f_0$ and $\sigma$ mesons was considered in Ref.
\cite{nutral}. In this paper, we consider only expressions in
regard to the interference with the background reactions.

As was shown in Refs. \cite{achasov-89,nutral}, the dominant
background is the $e^+e^-\to\phi\to\pi^0\rho\to\gamma\pi^0\pi^0$
reaction. The amplitude of the
$e^+e^-\to\rho\to\pi^0\omega\to\gamma\pi^0\pi^0$ reaction is much
less than the amplitude of the
$e^+e^-\to\phi\to\pi^0\rho\to\gamma\pi^0\pi^0$ reaction. In its
turn, the amplitude of the
$e^+e^-\to\omega\to\pi^0\rho\to\gamma\pi^0\pi^0$ reaction is much
less than the amplitude of the
$e^+e^-\to\rho\to\pi^0\omega\to\gamma\pi^0\pi^0$ reaction.

The amplitude of the background decay
$\phi(p)\to\pi^0\rho\to\gamma(q)\pi^0(k_1)\pi^0(k_2)$ is written
in the following way:
\begin{equation}
M_{back}=g_{\rho\pi^0\phi}g_{\rho\pi^0\gamma}\phi_{\alpha}p_{\nu}
\epsilon_{\delta}q_{\epsilon}\epsilon_{\alpha\beta\mu\nu}
\epsilon_{\beta\delta\omega\epsilon}\left
(\frac{k_{1\mu}k_{2\omega}}
{D_{\rho}(q+k_2)}+\frac{k_{2\mu}k_{1\omega}}{D_{\rho}(q+k_1)}\right
).
\end{equation}

The amplitude of the signal
$\phi\to\gamma(f_0+\sigma)\to\gamma\pi^0\pi^0$ takes into account
the mixing of $f_0$ and $\sigma$ mesons, see \cite{nutral},

\begin{equation}
M_{f_0}=g(m)e^{i\delta_B}\left ((\phi\epsilon)- \frac{(\phi
q)(\epsilon p)}{(pq)}\right )\left
(\sum_{R,R'}g_{RK^+K^-}G_{RR'}^{-1}g_{R'\pi^0\pi^0}\right )
\label{f0signal}\,,
\end{equation}
where $R,R'=f_0,\sigma$. The matrix of propagators is defined in
Ref. \cite{nutral}. The phase of the signal amplitude is formed by
the phase of the triangle diagram ($\phi\to K^+K^-\to\gamma R$)
and by the phase of $\pi\pi$ scattering which in its turn  is
defined by the phase of the $f_0-\sigma$ complex, and by the phase
of the elastic background of $\pi\pi$ scattering, $\delta_B$, see
details in Refs. \cite{phase,nutral,ach-84}.

The mass spectrum of the process is
\begin{equation}
\frac{\Gamma(\phi\to\gamma\pi^0\pi^0)}{dm}=\frac{d\Gamma_{f_0}(m)}{dm}+
\frac{d\Gamma_{back}(m)}{dm}\pm \frac{d\Gamma_{int}(m)}{dm},
\end{equation}
where the mass spectrum of the signal has the form
 \begin{equation}
\frac{d\Gamma_{f_0}(m)}{dm}=\frac{|g(m)|^2\sqrt{m^2-4m_{\pi}^2}(m_{\phi}^2-m^2)}
{3(4\pi)^3m_{\phi}^3}\left
|\sum_{R,R'}g_{RK^+K^-}G_{RR'}^{-1}g_{R'\pi^0\pi^0}\right |^2.
\label{f0}
\end{equation}

The mass spectrum for the background process
$e^+e^-\to\phi\to\pi^0\rho\to\gamma\pi^0\pi^0$ is
\begin{equation}
\frac{d\Gamma_{back}(m)}{dm}=\frac{1}{2}\frac{(m_{\phi}^2-m^2)\sqrt{m^2-4m_{\pi}^2}}
{256\pi^3m_{\phi}^3} \int_{-1}^{1}dxA_{back}(m,x)\,,
\label{phonf0}
\end{equation}
where

\begin{eqnarray}
&&A_{back}(m,x)=\frac{1}{3}\sum\left |M_{back}\right |^2=\nonumber \\
&&=\frac{1}{24}g_{\phi\rho\pi}^2g_{\rho\pi\gamma}^2 \left \{ \left
(m_{\pi}^8+2m^2m_{\pi}^4\tilde{m_{\rho}^2}-
4m_{\pi}^6\tilde{m_{\rho}^2}+2m^4\tilde{m_{\rho}^4}-\right.\right. \nonumber \\
&&\left.\left. 4m^2m_{\pi}^2\tilde{m_{\rho}^4}+
6m_{\pi}^4\tilde{m_{\rho}^4}+2m^2\tilde{m_{\rho}^6}-4m_{\pi}^2\tilde{m_{\rho}^6}+
\tilde{m_{\rho}^8}-2m_{\pi}^6m_{\phi}^2-\right.\right. \nonumber \\
&&\left.\left. 2m^2m_{\pi}^2\tilde{m_{\rho}^2}m_{\phi}^2+
2m_{\pi}^4\tilde{m_{\rho}^2}m_{\phi}^2-2m^2\tilde{m_{\rho}^4}m_{\phi}^2+
2m_{\pi}^2\tilde{m_{\rho}^4}m_{\phi}^2-2\tilde{m_{\rho}^6}m_{\phi}^2+\right.\right.
\nonumber
\\ &&\left.\left. m_{\pi}^4m_{\phi}^4+ \tilde{m_{\rho}^4}m_{\phi}^4\right)\left
(\frac{1}{|D_{\rho}(m_{\rho})|^2}+
\frac{1}{|D_{\rho}(\tilde{m_{\rho}^*})|^2}\right )+\left (m_{\phi}^2-m^2\right )\left (m^2-\right.\right. \nonumber\\
&&\left.\left. 2m_{\pi}^2+2\tilde{m_{\rho}^2}-m_{\phi}^2\right
)\left (2m^2m_{\pi}^2+2m_{\pi}^2m_{\phi}^2-m^4\right
)\frac{1}{|D_{\rho}(\tilde{m_{\rho}^*})|^2}+\right.
\nonumber \\
&&\left. 2Re\left
(\frac{1}{D_{\rho}(m_{\rho})D^*_{\rho}(\tilde{m_{\rho}^*}}\right )
\left (m_{\pi}^8-
m^6\tilde{m_{\rho}^2}+2m^4m_{\pi}^2\tilde{m_{\rho}^2}+\right.\right.
\nonumber \\
&&\left.\left.
2m^2m_{\pi}^4\tilde{m_{\rho}^2}-4m_{\pi}^6\tilde{m_{\rho}^2}-
4m^2m_{\pi}^2\tilde{m_{\rho}^4}+6m_{\pi}^4\tilde{m_{\rho}^4}+\right.\right.
\nonumber
\\ &&\left.\left. 2m^2\tilde{m_{\rho}^6}-4m_{\pi}^2\tilde{m_{\rho}^6}+\tilde{m_{\rho}^8}+
m^2m_{\pi}^4m_{\phi}^2-2m_{\pi}^6m_{\phi}^2+
2m^4\tilde{m_{\rho}^2}m_{\phi}^2-\right.\right. \nonumber \\
&&\left.\left. 4m^2m_{\pi}^2\tilde{m_{\rho}^2}m_{\phi}^2+
2m_{\pi}^4\tilde{m_{\rho}^2}m_{\phi}^2-m^2\tilde{m_{\rho}^4}m_{\phi}^2+
2m_{\pi}^2\tilde{m_{\rho}^4}m_{\phi}^2-2\tilde{m_{\rho}^6}m_{\phi}^2-\right.\right.
\nonumber
\\ &&\left.\left. m_{\pi}^4m_{\phi}^4-m^2\tilde{m_{\rho}^2}m_{\phi}^4+
2m_{\pi}^2\tilde{m_{\rho}^2}m_{\phi}^4+\tilde{m_{\rho}^4}m_{\phi}^4\right
) \right \}
\end{eqnarray}
and
\begin{eqnarray}
&&\tilde{m_{\rho}^2}=m_{\pi}^2+\frac{(m_{\phi}^2-m^2)}{2}
\left (1-x\sqrt{1-\frac{4m_{\pi}^2}{m^2}}\right )\nonumber \\
&&\tilde{m_{\rho}^{*2}}=m^2_{\phi}+2m_{\pi}^2-m^2-\tilde{m_{\rho}}^2\,.
\end{eqnarray}

The interference between the amplitudes of the background process
and the signal has the form
\begin{equation}
\frac{d\Gamma_{int}(m)}{dm}=\frac{1}{\sqrt{2}}\frac{\sqrt{m^2-4m_{\pi}^2}}
{256\pi^3m_{\phi}^3} \int_{-1}^{1}dxA_{int}(m,x)\,,
 \label{intf0}
\end{equation}
where
\begin{eqnarray}
&&A_{int}(m,x)=\frac{2}{3}Re\sum M_fM_{back}^*=\nonumber \\
&&\frac{1}{3}Re\left\{g(m)e^{\delta_b}g_{\phi\rho\pi}g_{\rho\pi^0\gamma}
(\sum_{R,R'}g_{RK^+K^-}G_{RR'}^{-1}g_{R'\pi^0\pi^0})
\left(\frac{(\tilde{m_{\rho}^2}-m_{\pi}^2)^2m_{\phi}^2-(m_{\phi}^2-m^2)^2\tilde{m_{\rho}^2}
}{D_{\rho}^*(\tilde{m_{\rho}})}+ \right.\right. \nonumber\\
&&\left.\left.
\frac{(\tilde{m_{\rho}^{*2}}-m_{\pi}^2)^2m_{\phi}^2-(m_{\phi}^2-m^2)^2
\tilde{m_{\rho}^{*2}}
}{D_{\rho}^*(\tilde{m_{\rho}}^*)}\right)\right\}\,.
\end{eqnarray}
 The factor $1/2$ in (\ref{phonf0}) and the factor  $1/\sqrt{2}$ in
(\ref{intf0}) take into account the identity of pions. In
(\ref{f0}), the identity of pions  is taken into account by the
definition of the coupling constant
$g_{R\pi^0\pi^0}=g_{R\pi^+\pi^-}/\sqrt{2}$. For the fitting of the
experimental data we use the model of $\pi\pi$ scattering
considered in Ref. \cite{nutral}. The phase of the elastic
background of $\pi\pi$ scattering is taken in the form
$\delta_B=b\sqrt{m^2-4m_{\pi}^2}$. We fit simultaneously the phase
of $\pi\pi$ scattering and the experimental data on the decay
$\phi\to\gamma\pi^0\pi^0$.

The fit of the experimental data \cite{snd-ivan}, obtained  using
the total statistics of SND detector, and the data on the $\pi\pi$
scattering phase
\cite{{hyams},{estabrook},{martin},{srinivasan},{rosselet}},
taking  the value $g_{\rho\pi^0\gamma}=0.295\pm0.037\ GeV^{-1}$
obtained from the data on the $\rho\to\pi^0\gamma$ decay
\cite{dolinsky} with the help of the following expression:
\begin{equation}
\Gamma(\rho\to\pi^0\gamma)=\frac{g_{\rho\pi^0\gamma}^2}{96\pi
m_{\rho}^3} \left (m_{\rho}^2-m_{\pi}^2\right )^3,
\end{equation}
gives the constructive interference and the following parameters:
\begin{eqnarray}
\label{snd}
 &&g_{f_0K^+K^-}=4.021\pm0.011\  \mbox{GeV},\ \
g_{f_0\pi^0\pi^0}=1.494\pm0.021\ \mbox{GeV},\ \
m_{f_0}=0.996\pm0.0013\ \mbox{GeV}, \nonumber
\\ && g_{\sigma K^+K^-}=0, \ \ g_{\sigma\pi^0\pi^0}=2.58\pm0.02\ \mbox{GeV}, \ \
m_{\sigma}=1.505\pm0.012\ \mbox{GeV}, \nonumber
\\ &&b=75\pm2.1\ (1^{\circ}/GeV),\ \ \  C=0.622\pm0.04\  \mbox{GeV}^2,\ \
g_{f_0K^+K^-}^2/4\pi=1.29\pm0.017\  \mbox{GeV}^2.
\end{eqnarray}

The constant  $C$ takes into account effectively the contribution
of multi particle intermediate states in the
$f_0\leftrightarrow\sigma$ transition in  $G_{RR'}$ matrix, see
Ref. \cite{nutral}, and incorporates  the subtraction constant for
the $R\to(0^-0^-)\to R'$ transition. We treat this constant as a
free parameter.

\begin{figure}
\centerline{\epsfxsize=12cm \epsfysize=8cm \epsfbox{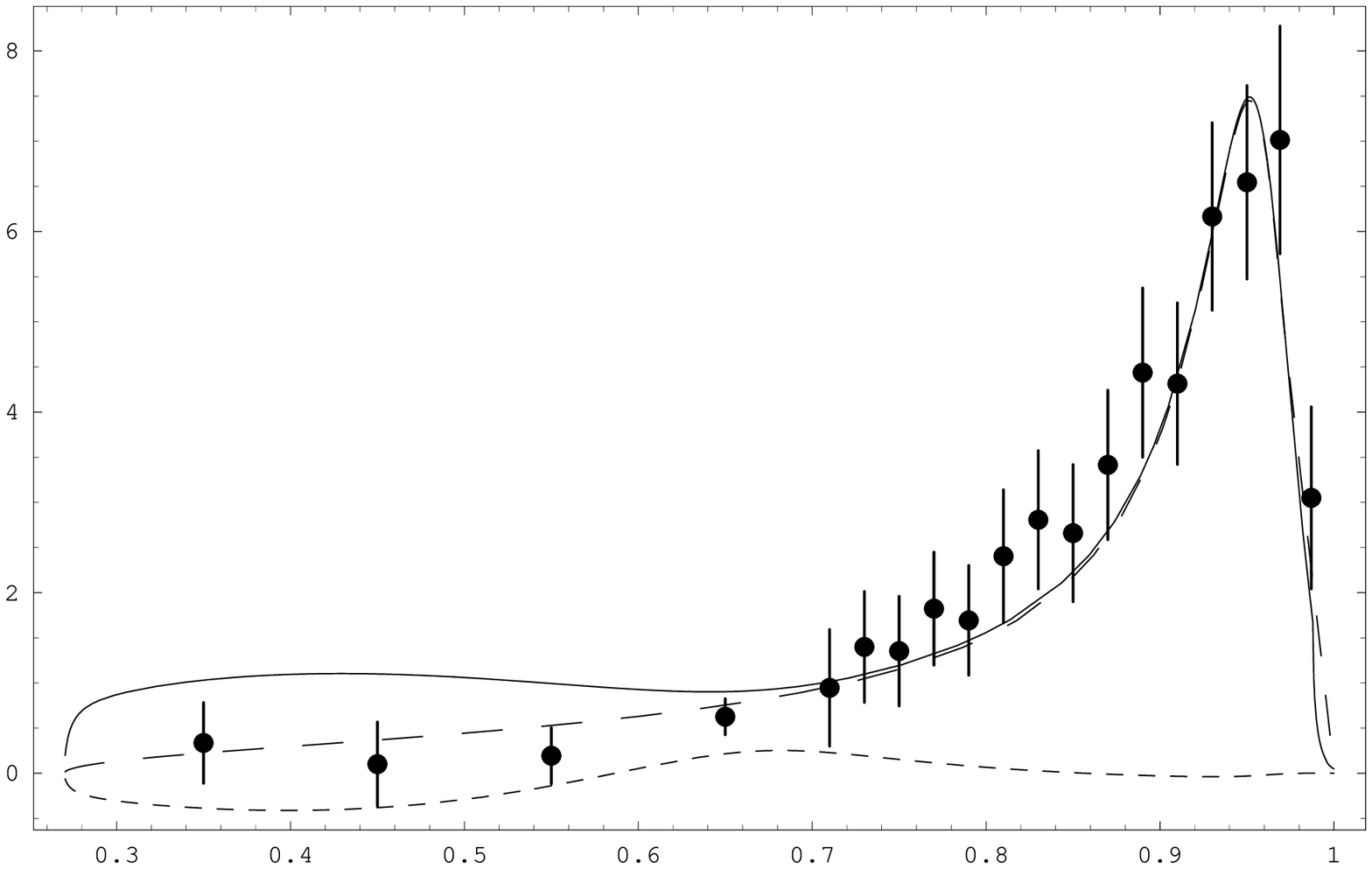}}
 \caption{
  Fitting of $dBR(\phi\to\gamma\pi^0\pi^0)/dm\times 10^{4}\mbox{GeV}^{-1}$  with
the background is shown with the solid line, the signal
contribution is shown with the dashed line. The dotted line is the
interference term. The data are from the SND detector.
  } \label{figf0snd}
\end{figure}

\begin{figure}
\centerline{\epsfxsize=12cm \epsfysize=8cm \epsfbox{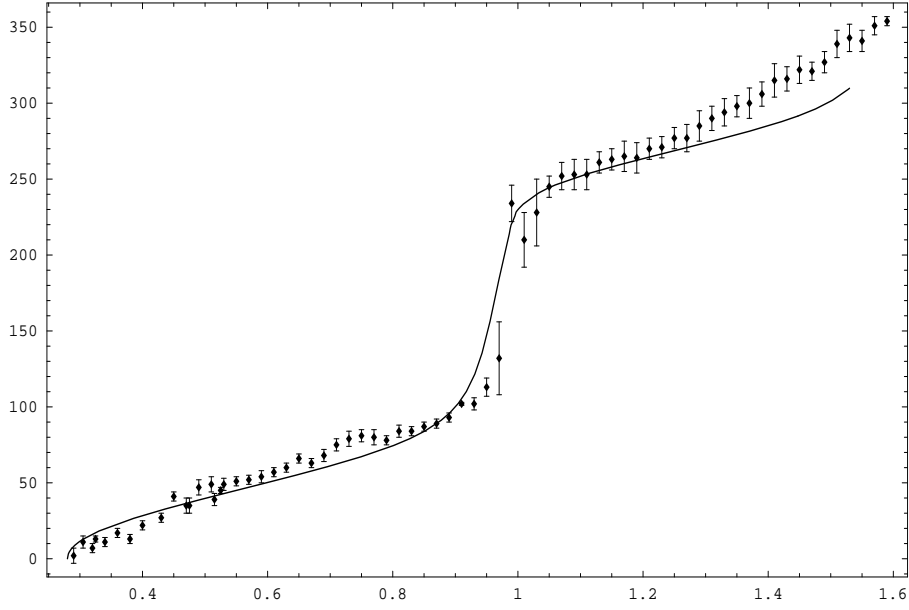}}
 \caption{ Fitting of the phase $\delta^0_0$ of
 $\pi\pi$ scattering.}
\label{figphasesnd}
\end{figure}

 The total branching ratio, with interference being taken into
 account, is
  $BR(\phi\to(\gamma
f_0+\pi^0\rho)\to\gamma\pi^0\pi^0)=(1.26\pm0.29)\cdot10^{-4}$, the
branching ratio of the signal is $BR(\phi\to\gamma
f_0\to\gamma\pi^0\pi^0)=(1.01\pm0.23)\cdot10^{-4}$,
 the branching ratio of the background is
$BR(\phi\to\rho^0\pi^0\to\gamma\pi^0\pi^0)=0.18\cdot10^{-4}$. The
 results of fitting are shown in  Figs. \ref{figf0snd} and
\ref{figphasesnd}.

Note, that for our aim, the phase  in the region $m_{\pi\pi}<1.1$
GeV is important.

The authors of Ref. \cite{snd-ivan} fit the data taking into
account the background reaction
$\phi\to\rho^0\pi^0\to\gamma\pi^0\pi^0$. The parameters found
\cite{snd-ivan} are $m_{f_0}=0.9698\pm0.0045$, $g_{f_0K^+
K^-}^2/4\pi=2.47\pm^{0.73}_{0.51}$ GeV$^2$ and $g_{f_0\pi^+\pi^-}
^2/4\pi=0.54\pm^{0.09}_{0.08}$ GeV$^2$. They are different from
the parameters found in our fitting. The difference is due to the
fact that we perform the simultaneous fitting of the data on the
decay $\phi\to\gamma\pi^0\pi^0$ and the data on the S-wave phase
of $\pi\pi$ scattering, taking into account the mixing of  $f_0$
and $\sigma$ mesons.

In addition, in Ref. \cite{snd-ivan}, the interference between the
background and signal is found from the fitting meanwhile in our
paper the interference is calculated. The branching ratio of the
background
$BR(\phi\to\rho^0\pi^0\to\gamma\pi^0\pi^0)=0.12\cdot10^{-4}$ used
in Ref. \cite{snd-ivan} is taken from Ref. \cite{bramon} in which
the coupling constant $g_{\rho^0\pi^0\gamma}$ is less by 25\% than
resulting from the experiment. In our paper, the background is
calculated on the basis of experiment and is accordingly larger,
 $BR(\phi\to\rho^0\pi^0\to\gamma\pi^0\pi^0)=0.18\cdot10^{-4}$.
 Note that in Ref. \cite{snd-ivan}, in contrast to us, the fitting
 is performed  taking into account  the event distribution inside
 each bin.

The fitting of the experimental data of the CMD-2 detector
\cite{cmd} and the data on the $\pi\pi$ scattering phase
\cite{{hyams},{estabrook},{martin},{srinivasan},{rosselet}} gives
the constructive interference and the following parameters:

\begin{eqnarray}
\label{cmd2}
 &&g_{f_0K^+K^-}=3.874\pm0.17\ \mbox{GeV},\ \
g_{f_0\pi^0\pi^0}=0.536\pm0.03\ \mbox{GeV},\ \
m_{f_0}=1.0019\pm0.002\ \mbox{GeV}, \nonumber \\ && g_{\sigma
K^+K^-}=0, \ \ g_{\sigma\pi^0\pi^0}=2.61\pm0.1\ \mbox{GeV}, \ \
m_{\sigma}=1.585\pm0.015\ \mbox{GeV}, \nonumber
\\ &&b=70.7\pm2.0\ (1^{\circ}/GeV),\ \ \  C=-0.593\pm0.06\
\mbox{GeV}^2,\ \ g_{f_0K^+K^-}^2/4\pi=1.19\pm0.03\  \mbox{GeV}^2.
\end{eqnarray}

\begin{figure}
\centerline{\epsfxsize=12cm \epsfysize=8cm \epsfbox{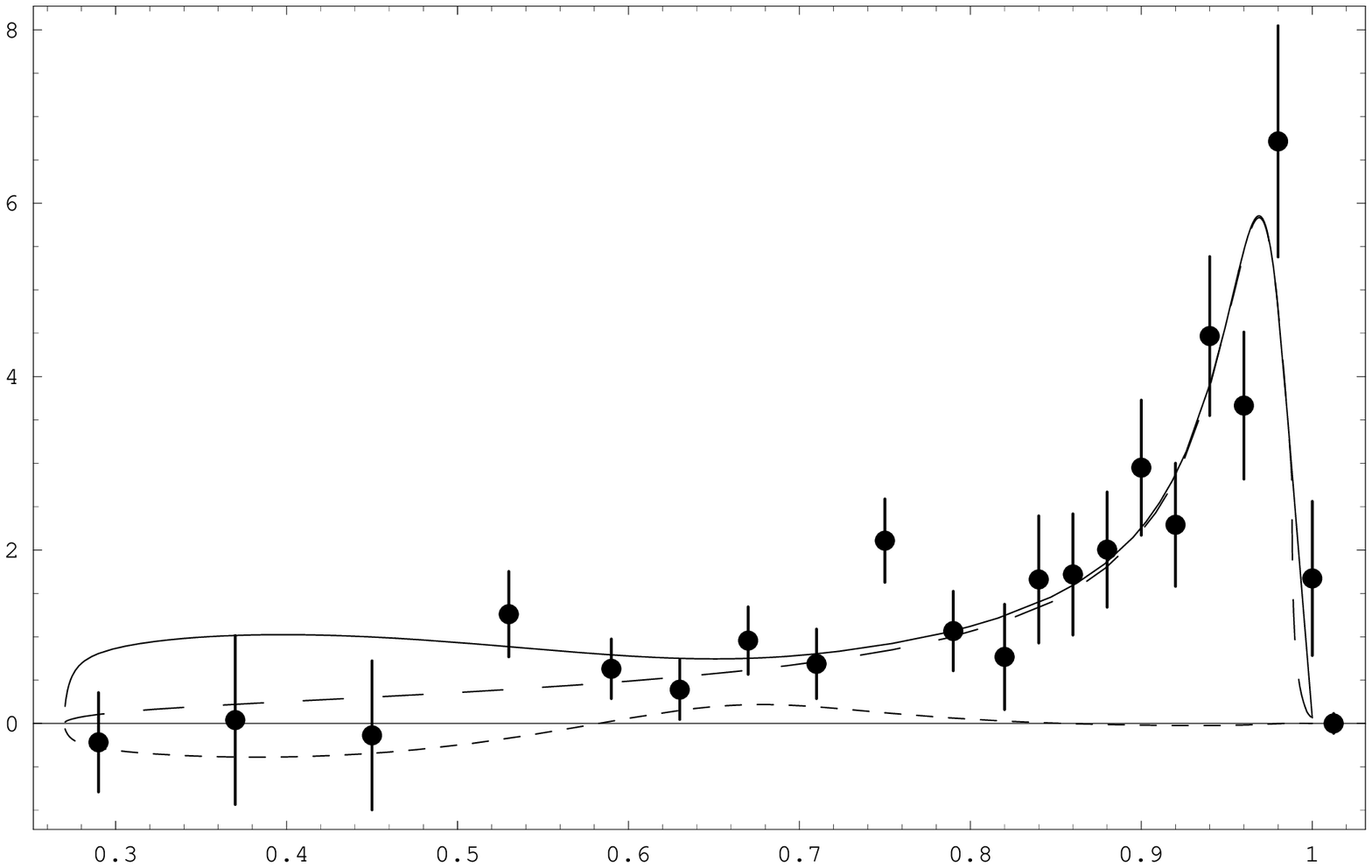}}
 \caption{ Fitting of $dBR(\phi\to\gamma\pi^0\pi^0)/dm\times 10^{4}\mbox{GeV}^{-1}$  with
the background is shown with the solid line, the signal
contribution is shown with the dashed line.  The dotted line is
the interference term. The data are from the CMD-2 detector.
 } \label{figf0cmd}
\end{figure}

\begin{figure}
\centerline{\epsfxsize=12cm \epsfysize=8cm \epsfbox{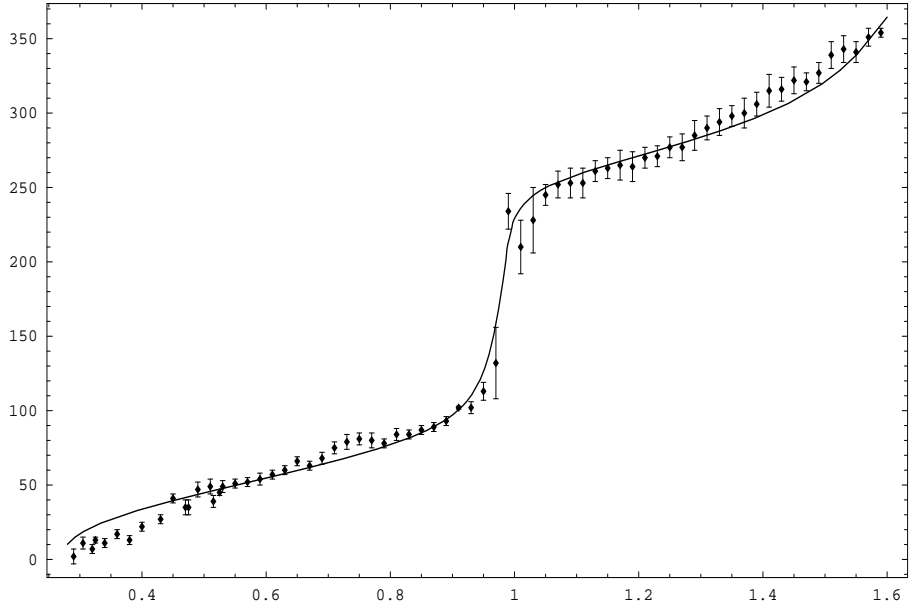}}
 \caption{ Fitting of the phase $\delta^0_0$ of
 $\pi\pi$ scattering.}
\label{figphasecmd}
\end{figure}

 The total branching ratio, taking into account the interference, is  $BR(\phi\to(\gamma
f_0+\pi^0\rho)\to\gamma\pi^0\pi^0)=(0.98\pm0.21)\cdot10^{-4}$, the
branching ratio of the signal is $BR(\phi\to\gamma
f_0\to\gamma\pi^0\pi^0)=(0.74\pm0.2)\cdot10^{-4}$,
 the branching ratio of the background is
$BR(\phi\to\rho^0\pi^0\to\gamma\pi^0\pi^0)=0.18\cdot10^{-4}$. The
 results of fitting are shown in  Figs. \ref{figf0cmd} and
\ref{figphasecmd}.


The parameters found in  \cite{cmd}, which are
$m_{f_0}=0.969\pm0.005$, $g_{f_0 K^+ K^-}^2/4\pi=1.49\pm0.36\
GeV^2$ and $g_{f_0\pi^+\pi^-} ^2/4\pi=0.4\pm0.06\ GeV^2$, are
different from  the parameters found in our fitting. The
difference is due to the fact that we perform the simultaneous
fitting of the data on the decay $\phi\to\gamma\pi^0\pi^0$ and the
data on the S-wave phase of the $\pi\pi$ scattering, taking into
account the mixing of  $f_0$ and $\sigma$ mesons and taking into
account the background reaction
$\phi\to\rho^0\pi^0\to\gamma\pi^0\pi^0$.

One can see from  Figs.  \ref{figf0snd} and \ref{figf0cmd} that
the influence of the background process on the spectrum of the
$\phi\to\gamma\pi^0\pi^0$ decay  is negligible in the wide region
of the $\pi^0\pi^0$ invariant mass, $m_{\pi\pi}>670$ MeV, or when
photon energy less than $300$ MeV.

In the meantime, the difference from the experimental data is
observed in the region $m_{\pi\pi}<670$ MeV. We suppose this
difference is due to the fact that in the experimental processing
of the $e^+e^-\to\gamma\pi^0\pi^0$ events the background events
$e^+e^-\to\omega\pi^0\to\gamma\pi^0\pi^0$ are excluded with the
help of the invariant mass cutting and simulation, in so doing the
part of the $e^+e^-\to\phi\to\rho\pi^0\to\gamma\pi^0\pi^0$ events
is excluded as well.

It should be noted that the SND and CMD-2 data on the branching
ratios of the $\phi\to\gamma\pi^0\pi^0$ decay are quite
consistent, in the meantime, the SND and CMD-2 data on the shapes
of the spectra of  the $\pi^0\pi^0$ invariant mass are rather
different. The CMD-2 shape is noticeably more narrow, compare
Figs. \ref{figf0snd} and \ref{figf0cmd}. This difference reflects
on the coupling constant $g_{f_0\pi^0\pi^0}$ and the constant $C$,
which are quite different, see Eqs. (\ref{snd}) and (\ref{cmd2}).
In all probability, this difference will disappear when the CMD-2
group processes the total statistics.

\section{Conclusion.}

The experimental data give evidence not only in favor of the
four-quark model but in favor of the dynamical model suggested in
Ref. \cite{achasov-89}, in which the discussed decays proceed
through the kaon loop, $\phi\to K^+ K^- \to\gamma f_0(a_0)$.

\begin{figure}
\centerline{\epsfxsize=10cm \epsfysize=6cm \epsfbox{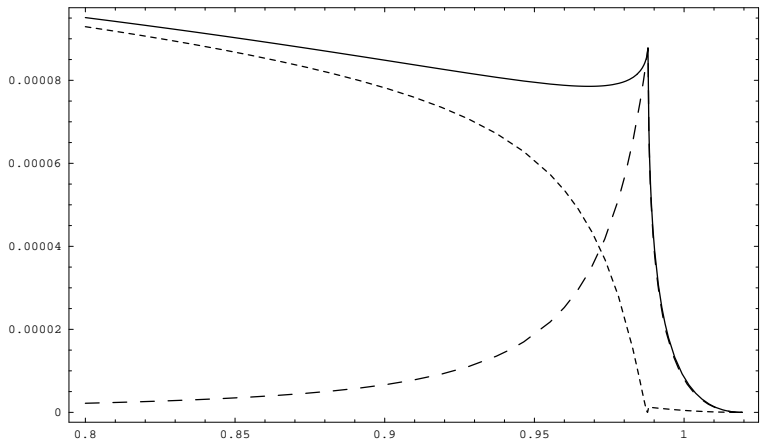}}
 \caption{ The function $|g(m)|^2$ is drawn with the solid line. The contribution of the
  imaginary part is drawn with the dashed line. The contribution of the real part
   is drawn with the dotted line.} \label{g}
\end{figure}

Indeed, according to the gauge invariance condition, the
transition amplitude $\phi\to\gamma f_0(a_0)$ is proportional to
the electromagnetic tensor $F_{\mu\nu}$ (in our case to the
electric field). Since there are no pole terms in our case, the
function $g(m)$ in (\ref{a0signal}) and (\ref{f0signal}) is
proportional to the energy of photon
$\omega=(m_{\phi}^2-m^2)/2m_{\phi}$ in the soft photon region. To
describe the experimental spectra, the function $|g(m)|^2$ should
be smooth (almost constant) in the range  $m\leq0.99$ GeV, see
Eqs. (\ref{spectruma0}) and (\ref{f0}). Stopping the function
$\omega^2$ at $\omega_0=30$ MeV, using the form-factor of the form
$1/(1+R^2\omega^2)$, requires $R\approx 100$ GeV$^{-1}$. It seems
to be incredible to explain the formation of such a huge radius in
hadron physics. Based on the large, by hadron physics standard,
$R\approx10$ GeV$^{-1}$, one can obtain an effective maximum of
the mass spectra under discussion only near 900 MeV. In the
meantime, the $K^+K^-$ loop  gives the natural description to this
threshold effect, see Fig. \ref{g}.

\begin{figure}
\centerline{\epsfxsize=10cm \epsfysize=6cm \epsfbox{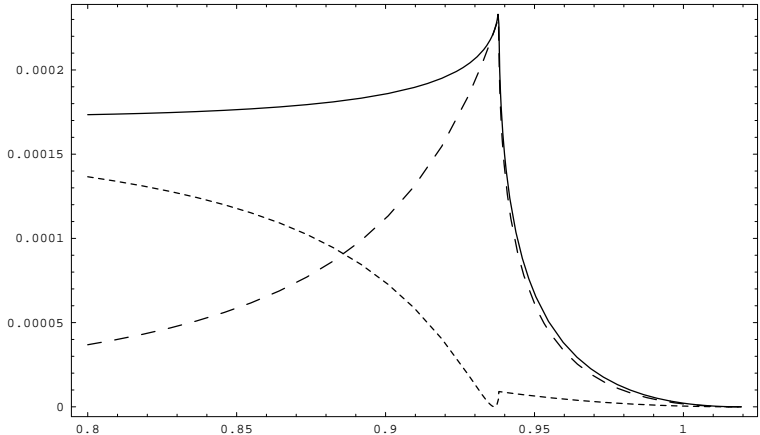}}
 \caption{ The function $|g(m)|^2$ for  $m_K=469$ MeV is drawn
  with the solid line. The contribution of the imaginary  is drawn with the dashed line.
  The contribution of the real part is drawn with the dotted line.} \label{gg}
\end{figure}

 To demonstrate the threshold character of this effect we present
 Fig. \ref{gg} in which the function $|g(m)|^2$ is shown in the
case of $K^+$ meson mass is 25 MeV less than in reality. One can
see that in the region 950-1020 MeV the function $|g(m)|^2$ is
suppressed by the $\omega^2$ low. In the mass spectrum this
suppression is increased by one more power of $\omega$, see Eqs.
(\ref{spectruma0}) and (\ref{f0}), so that we cannot see the
resonance in the region 980-995 MeV. The maximum in the spectrum
is effectively shifted to the region 935-950 MeV. In truth this
means that $a_0(980)$ and $f_0(980)$ resonances are seen in the
radiative decays of $\phi$ meson owing to the $K^+K^-$
intermediate state, otherwise the maxima in the spectra would be
shifted to 900 MeV.

It is worth noting that the $K^+K^-$ loop  model is practically
accepted by theorists, compare, for example, Ref. \cite{marco}
with Ref. \cite{br}, true there is exception \cite{anis}.

 It was noted already in paper \cite{achasov-89} that the imaginary
part of the $K^+K^-$  loop is calculated practically in a model
independent way making use of the coupling constants $g_{\phi
K^+K^-}$ and $g_{a_0(f_0)K^+K^-}$ due to the Low's theorem
\cite{low} for the photons with energy $\omega<100$ MeV which is
soft by the standard of strong interaction. In the same paper it
was noted that the real part of the loop  (with accuracy up to
20\% in the width of the $\phi\to\gamma f_0(a_0)$ decay) is
practically not different for the point-like particle and the
compact hadron with form-factor which has the cutting radius in
the momentum space about the mass of  $\rho$ meson
($m_{\rho}=0.77$ GeV). In contrast to the four-quark state which
is the compact hadron \cite{jaffe}, the bound $K\bar K$ state is
the extended state with the spatial radius
$R\sim1/\sqrt{m_K\epsilon}$, where $\epsilon$ is the binding
energy. Corresponding form-factor in the momentum space has the
radius of the order of $\sqrt{m_K\epsilon}\approx100$ MeV for
$\epsilon=20$ MeV, \cite{markushin}. The more detailed calculation
\cite{close-93} gives for the radius in the momentum space the
value $p_0=140$ MeV. As a result, the contribution of the virtual
intermediate $K^+K^-$ states in the $K^+K^-$ loop is suppressed by
the momentum distribution in the molecule, and the real part of
the loop amplitude is negligible \cite{shevchenko}. It leads to
the branching ratio much less  than the experimental one, as it
was noted above. In addition, the spectrum is much narrower in the
$K\bar K$ molecule case  that contradicts to the experiment, see
the behavior of the imaginary part contribution in Fig. \ref{g}
and in corresponding figures in \cite{shevchenko}.

 Of course,  the two-quark state is as compact as four-quark one.
The question arises, why is the branching ratio in the two-quark
model suppressed in comparison with the branching ratio in the
four-quark model? There are two reasons. First, the coupling
constant of two-quark states with the $K\bar K$ channel is
noticeably less \cite{nutral,ach-84} and, second, there is the
Okubo-Zweig-Iizuka (OZI) rule that is more important really.

If the isovector $a_0(980)$ meson is the two-quark state,  it has
no strange quarks. Hence \cite{achasov-89,nutral,ach-98}, the
decay $\phi\to\gamma a_0$ should be suppressed according to the
OZI rule. On the intermediate state level, the OZI rule is
formulated as compensation of the different intermediate states
\cite{lipkin,geiger,ach-kozh}. In our case these states are $K\bar
K$, $K\bar K^*+\bar KK^*$, $K^*\bar K^*$ and so on. Since, due to
the kinematical reason, the real intermediate state is the only
 $K^+K^-$ state, the compensation in the imaginary part is
 impossible and it destroys the OZI rule. The compensation should
 be in the real part of the amplitude only. As a result, the $\phi\to\gamma a_0$ decay
  in the two-quark  model is mainly due to the imaginary part of the
 amplitude and is much less intensive than in the four-quark
 model \cite{achasov-89,nutral}. In addition, in the two-quark model, $a_0(980)$ meson should
 appear in the  $\phi\to\gamma a_0$ decay as a noticeably more narrow
 resonance than in other processes, see the
behavior of the imaginary part contribution in Fig. \ref{g}.

 As regards to the isoscalar $f_0(980)$ state, there are two
possibilities  in the two-quark model. If $f_0(980)$ meson does
not contain the strange quarks  the all above mentioned arguments
about suppression of the  $\phi\to\gamma a_0$ decay and the
spectrum shape  are also valid for the  $\phi\to\gamma f_0$ decay.
Generally speaking, there could be the strong OZI violation for
the isoscalar $q\bar q$ states ( mixing of the $u\bar u$, $d\bar
d$ and $s\bar s$ states) with regard to the strong mixing of the
quark and gluon degree of freedom which is due to the
nonperturbative effects of QCD \cite{vainshtein}. But, an almost
exact degeneration of the masses of the isoscalar $f_0(980)$ and
isovector $a_0(980)$ mesons excludes  such a possibility. Note
also, the experiment points directly to the weak coupling of
$f_0(980)$ meson with gluons, $B(J/\psi\to\gamma
f_0\to\gamma\pi\pi)<1.4\cdot10^{-5}$ \cite{eigen}.

If $f_0(980)$ meson is close to the $s\bar s$ state
\cite{ach-98,tornqvist} there is  no suppression due to the the
OZI rule. Nevertheless, if $a_0(980)$ and $f_0(980)$ mesons are
the members of the same multiplet, the $\phi\to\gamma f_0$
branching ratio,
$BR(\phi\to\gamma\pi^0\pi^0)=(1/3)BR(\phi\to\gamma\pi\pi)
 \approx1.8\cdot10^{-5}$,
  is  significantly less than that in the four-quark
model,  due to the relation between the coupling constants with
the $K\bar K$, $\pi\eta$ and $K\bar K$, $\pi\pi$ channels
inherited in the two-quark model, see Refs.
\cite{achasov-89,nutral}. In addition, in this case there is no
natural explanation of the $a_0$ and $f_0$ mass degeneration.

Only in the case when the nature of $f_0(980)$ meson in no way
related to the nature of $a_0(980)$ meson (which, for example, is
the four-quark state) the branching ratio experimentally observed
$\phi\to\gamma f_0$ could be explained   by $s\bar s$ nature of
$f_0(980)$ meson. But, from the theoretical point of view, such a
possibility seems awful  \cite{ach-98}.

\section{Acknowledgement}

 We gratefully acknowledge discussions
with V.P. Druzhinin, A.A. Kozhevnikov, G.N. Shestakov,  and Z.K.
Silagadze. This work was supported in part by INTAS-RFBR, grant
IR-97-232.


\begin{thebibliography}{99}
\bibitem{markushin}
Unfortunately, in the interesting paper V.E. Markushin, Eur.
Phys.J., A8, 389 (2000), the potential in the momentum space was
taken as the momentum distribution in the molecule instead of the
wave function in the momentum space. But the momentum distribution
radius of the potential is  5-8 times as large as one of the wave
function, that was the reason for the misleading conclusion on the
possibility to explain the Novosibirsk reasults in the molecule
case.
\end{thebibliography}
\end{document}